\documentclass[lettersize,journal]{IEEEtran}
\usepackage{amsmath,amsfonts}
\usepackage{array}
\usepackage[caption=false,font=normalsize,labelfont=sf,textfont=sf]{subfig}
\usepackage{textcomp}
\usepackage{stfloats}
\usepackage{url}
\usepackage{verbatim}
\usepackage{graphicx}
\usepackage{cite}
\usepackage{tabularx}
\usepackage{multirow}
\usepackage{makecell}
\usepackage{calligra}
\usepackage{algpseudocode} 
\usepackage{amsmath} 
\usepackage{etoolbox}
\usepackage{threeparttable}
\usepackage{diagbox}
\usepackage{amssymb}
\usepackage{graphicx}
\usepackage[linesnumbered, ruled]{algorithm2e}
\usepackage[table,xcdraw]{xcolor}

\begin{document}

\title{StegGuard: Fingerprinting Self-supervised Pre-trained Encoders via Secrets Embeder and Extractor}

\author{Xingdong Ren, Tianxing Zhang, Hanzhou Wu, Xinpeng Zhang, Yinggui Wang, Guangling Sun* ~\IEEEmembership{}}
\maketitle

\begin{abstract}
In this work, we propose StegGuard, a novel fingerprinting mechanism to verify the ownership of the suspect pre-trained encoder using steganography. A critical perspective in StegGuard is that the unique characteristic of the transformation from an image to an embedding, conducted by the pre-trained encoder, can be equivalently exposed how an embeder embeds secrets into images and how an extractor extracts the secrets from encoder's embeddings with a tolerable error after the secrets are subjected to the encoder’s transformation. While each independent encoder has a distinct transformation, the piracy encoder has a similar transformation to the victim. Based on these, we learn a pair of secrets embeder and extractor as the fingerprint for the victim encoder. We introduce a frequency-domain channel attention embedding block into the embeder to adaptively embed secrets into suitable frequency bands. During verification, if the secrets embedded into the query images can be extracted with an acceptable error from the suspect encoder's embeddings, the suspect encoder is determined as piracy; otherwise independent. Extensive experiments demonstrate that depending on a very limited number of query images, StegGuard can reliably identify across varied independent encoders, and is robust against model stealing related attacks including model extraction, fine-tuning, pruning, embedding noising and shuffle.
\end{abstract}


\section{Introduction}

\IEEEPARstart{A}{s} a typical representation learning paradigm, self-supervised learning(SSL) \cite{paper21,paper22,paper23} has been popularly used to pre-train encoders. Embedding, as output of learned pre-trained encoder, can be leveraged in various downstream tasks. Since the SSL requires a huge number of training data and consumes heavy computation resources, the well-trained pre-trained encoder has tended to be achieved by tech giants and is thereafter deployed on the cloud platform and offered as a paid service to users. This trend has become a new business pattern known as Encoder as a Service(EaaS).\footnote{\url{https://openai.com/api/}}$^,$\footnote{\url{https://www.clarifai.com/}} However, it has been a public concern that pre-trained encoders deployed on cloud platforms may face serious threats from potential adversaries as shown in Fig. 1. Legitimate users will pay for the encoder and then develop them into different downstream models. Conversely, the adversary steals the pre-trained encoders through model extraction attack and earns illegal profits from the stolen encoders, and even performs further attacks such as backdoor attacks and adversarial example attacks, which will spur serious risks to the application of EaaS. Therefore, necessary defenses are needed for the encoders to safeguard the intellectual property of service providers.

\begin{figure}[!t]
\begin{center}
\includegraphics[width=\linewidth]{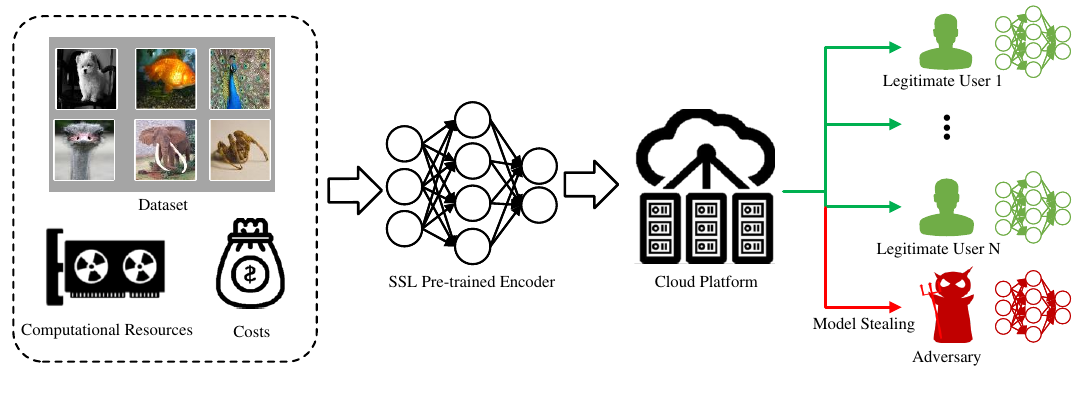}
\caption{A self-supervised pre-trained encoder is deployed on a cloud platform, and legitimate users train downstream models based on it, while the adversary illegally steals the encoder.}
\end{center}
\end{figure}

\IEEEpubidadjcol
The current mainstream model copyright protection techniques are categorized into model watermarking \cite{paper6,paper7,paper8,paper9} and fingerprinting \cite{paper10,paper11,paper12}. The watermarking strategy needs the owner to embed a watermark into the model. If the watermark can be extracted accurately from the suspect model, the model's copyright will be verified. Furthermore, model watermarking strategy can roughly fall into \emph{white-box watermarking} and \emph{black-box watermarking}. The former explicitly embeds the watermark into the internal parameters \cite{paper13}, probabilistic distribution of the weights \cite{paper14} or network structure \cite{paper15} of the model, but it requires white-box access to the suspect model during verification. The latter implicitly embeds the watermark through backdoor \cite{paper16,paper17} or adversarial example \cite{paper18} and solely requires the output of suspect model during verification, which is more suitable for the most realistic scenarios. On the other hand, model fingerprinting techniques verify model ownership totally depending on itself unique characteristic of individual model, which are figuratively referred to as ``model fingerprints". It can be in the form of an adversarial example \cite{paper19}, which describes the model decision boundary, or an image-label pair \cite{paper10,paper20} distributed on both sides of the decision boundary. This strategy results in most of the fingerprinting techniques that are limited in end-to-end classification models. However, with the popularity of SSL pre-trained encoders, the fingerprint technique demands to be applied into more situations. 

In this work, we propose StegGuard, a novel fingerprint mechanism for protecting IP of pre-trained encoder by using steganography. First, we put pre-trained encoder between a secrets embeder and an extractor. A key observation in StegGuard is that how embed secrets into images and how to extract the secrets from embeddings within an error threshold can be viewed as an equivalent characterization for transformation function of the encoder. Under this perspective, there exist different embeder and extractor for different encoder. Besides, due to the piracy's dependency on victim and separate learning process between an independent one and victim, the transformation function of victim is more consistent with piracy than the independent one. In a word, the learned secrets embeder and extractor can reasonably become fingerprint of each encoder and reliably used to discern the privacy encoder.  

However, it is intractable to learn the fingerprint for each suspect encoder, and how to attain highly performant fingerprint is the other vital problem. To address the first issue, we propose a fingerprinting scheme which is fulfilled by matching the fingerprint of victim with the suspect encoder. Obviously, the matching degree can be used as an evidence to determine whether or not the suspect encoder is piracy. To cope with the second issue, Frequency-domain channel attention Embedding block(FcaEmb) is introduced into the secrets embeder aiming at adaptively selecting suitable frequency bands to embed the secrets. Experiments show that StegGuard can reliably verify the ownership of the suspect encoder with a very limited number of query images.

\begin{itemize}
    \item We propose StegGuard, a novel fingerprint framework for protecting the intellectual property of SSL pre-trained encoder, in which the encoder connects secrets embeder and extractor. The embeder and extractor are learned concurrently while the encoder is frozen. As fingerprint of victim encoder, the embeder and extractor are used to verify the ownership of suspect encoder with a very limited number of query images.
    \item We introduce a Frequency-domain channel attention Embedding block (FcaEmb) into the secrets embeder. FcaEmb enables the embeder to adaptively choose suitable frequency bands to embed the secret into the image, thereby reducing the average error of secrets extracted by the extractor from embeddings and improving the visual invisibility of stego images. 
     \item Experimental results demonstrate that StegGuard can reliably identify across varied independent encoders, and is robust against model stealing related attacks including model extraction, fine-tuning, pruning, embedding noising and shuffle.
\end{itemize}

The rest of the paper is organized as follows: We describe the existing watermarking and fingerprinting methods and their limitations in Section II. Section III presents the motivation and methodology of StegGuard. Section IV evaluates and analyzes the effectiveness and robustness of StegGuard. Finally, we conclude this paper in Section V.

\section{Preliminaries and Related Works}
\subsection{Self-supervised Learning}
Self-supervised learning(SSL), as an emerging deep learning paradigm, is definitely employed to obtain high-performance pre-trained encoders by learning large amounts of unlabeled data. Since the process of SSL requires large amounts of data and consumes heavy computational resources, the popular trend is that tech giants train the high-performance SSL pre-trained encoders and then deploy them on cloud platforms and provide paid services to regular consumers, who can develop their desired downstream models based on the output embedding of the pre-trained encoders. Two main lines of SSL algorithms are currently used: contrast learning and masked autoencoding. We have selected three representative algorithms for each of the two lines in this work, SimCLR \cite{paper21}, MoCo v2 \cite{paper22} for the former, and MAE \cite{paper23} for the latter.

\subsection{Model Extraction Attack}
As a remarkable AI paradigm, SSL has attracted growing attention for its security and intellectual property protection. Some malicious users attempt to threaten the pre-trained encoder through illegal means, including membership inference attacks \cite{paper30,paper31}, backdoor attacks \cite{paper17}, etc. Currently, the most serious security threat to SSL pre-trained encoders is model extraction attack \cite{paper32,paper33,paper34}. In order to obtain a piracy encoder with similar performance to the victim encoder with black-box access, the adversary will adopt a set of unlabeled query images, then query the victim encoder, and the returned embedding will be used as the label to train the piracy encoder with supervised learning. Compared to pre-train the encoder from scratch, model stealing attack relies on the victim encoder so requires only a tiny amount of training images and lightweight computational resources to obtain the piracy encoder that performs similarly to the victim encoder.

\subsection{Copyright Protection of Deep Neural Network}

\subsubsection{Watermark}
Watermarking, as one of the most representative strategy for protecting copyright of Deep Neural Network(DNN), has been rapidly developed in recent years. Watermarking DNN has two major phases, watermark embedding and watermark verification. During watermark verification, there are two scenarios for the copyright owner: black-box and white-box. Wang \cite{paper13} et al. first achieve the white-box watermarking by embedding bit secrets into the parameters of the model. After that, Rouhani \cite{paper14} et al. further embed the watermark into the probabilistic distribution of the weights of the model and Zhao \cite{paper15} et al. directly embed the watermark into the model structure, which improves the efficiency and robustness of the watermark. However, in real-world scenarios, copyright owners often cannot directly access the internal weights of suspect models, so Adi \cite{paper24} et al. first use the backdoor technique \cite{paper16} to achieve black-box watermark embedding and verification, and Merrer \cite{paper18} et al. also achieve black-box watermarking by generating a set of adversarial examples. With the popularity of SSL pre-trained encoders, encoder watermarking techniques have also been developed. Wu \cite{paper25} et al. embed a watermark into the SSL pre-trained encoder through backdoor technique, and Cong \cite{paper26} et al. improve their watermark resistance to model extraction attacks.

\subsubsection{Fingerprint}
Model fingerprinting is a copyright protection technique that does not need to modify the parameters of the model. It relies on finding intrinsic characteristics as the fingerprint to identify the latent piracy model. Cao \cite{paper10} et al. generate adversarial examples near the decision boundary as the fingerprint, but it has a deficiency in robustness and stability. Lukas \cite{paper12} et al. and Zhao \cite{paper27} et al. use the transferability of adversarial examples as the fingerprint. Lukas et al. generate a set of transferable adversarial examples through the basic iteration method (BIM), which can be used to verify the piracy model. Based on this, Zhao et al. generate adversarial examples with similar model predictions to randomly selected images in the target class to further enhance the fingerprints' efficacy. However, these data point fingerprint strategies are incapable of effectively identifying the independent models. In addition, generating adversarial examples requires training a large number of surrogate models. Recently, Peng \cite{paper28} et al. introduce universal adversarial perturbation (UAP) to generate adversarial examples to enhance the robustness and effectiveness of fingerprinting. Meanwhile, they adopt supervised contrastive loss to distinguish independent models, but it also limits the application scope due to its computational cost. Furthermore, most existing model fingerprinting techniques are designed for classification models rather than the pre-trained encoder. Adam \cite{paper29} et al. use dataset inference to protect the intellectual property of the SSL pre-trained encoders. Differently, StegGuard simplifies the verification of the encoder's copyright by further extracting the output embeddings of the suspect encoder and translating them into understandable secrets for verification.

\textbf{\begin{figure}[!ht]
\begin{center}
\includegraphics[width=\linewidth]{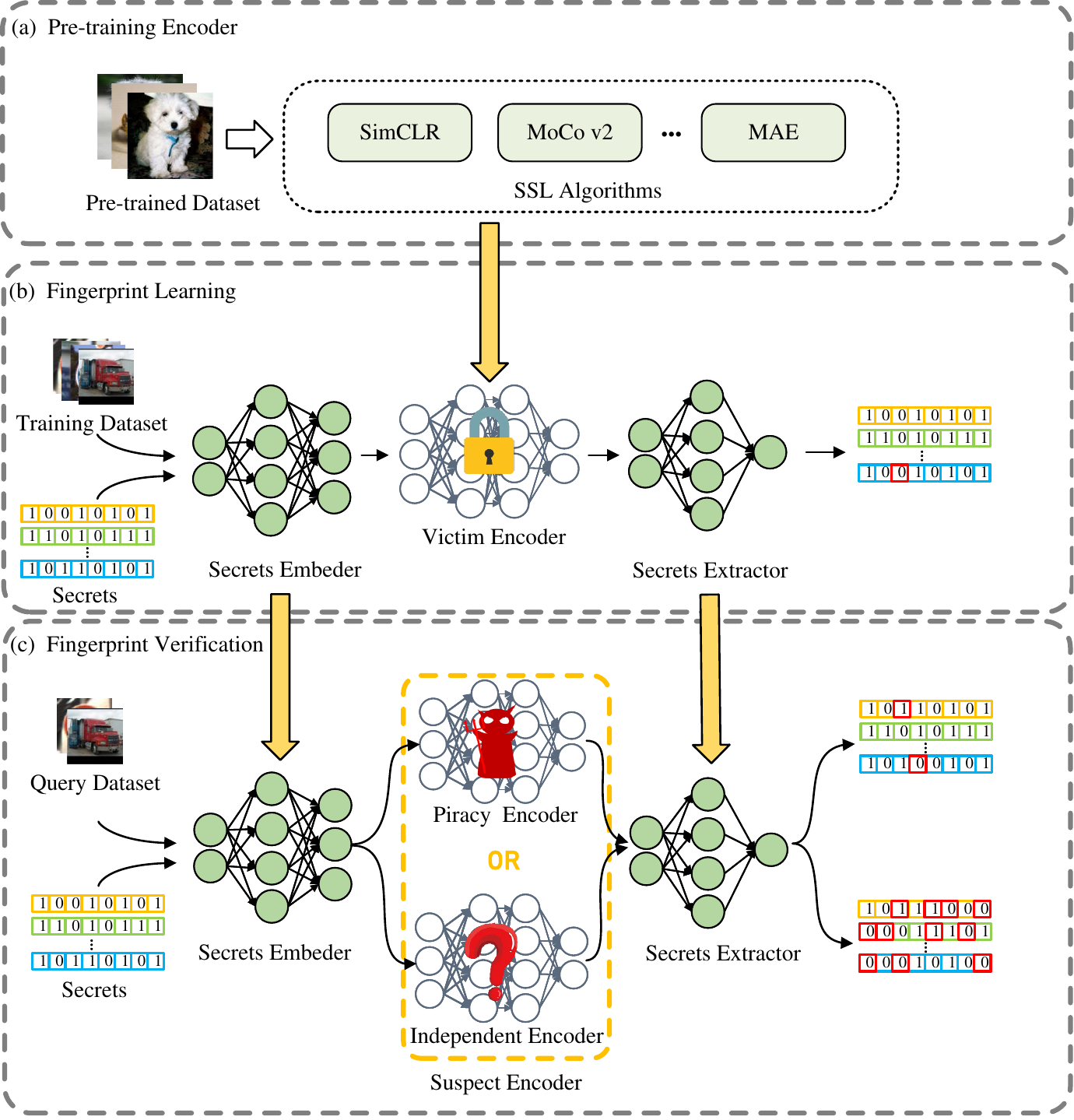}
\caption{Overview of StegGuard. (a) Pre-training encoder. (b) Fingerprint learning. (c) Fingerprint verification. The bit highlighted by red rectangle  denotes falsely extracted bit.}
\vspace{-0.7cm}
\end{center}
\end{figure}}

\section{Methodology}

\subsection{Threat Model}

The threat model considered in this work involves two parties: the \emph{adversary} and the \emph{defender}. The defender is the owner of the victim encoder, who deploys the encoder on the cloud platform and needs to protect the intellectual property of the encoder. Conversely, the adversary attempts to steal the victim encoder directly through the insider threat or indirectly through the model extraction attack.

\subsubsection{Adversary’s Capability and Knowledge} For the adversary, there are two ways to steal the victim encoder. The first is to directly obtain the complete encoder network and weights through insider threat, and the adversary will correspondingly fine-tune or prune the encoder to avoid the verification of the defender. The second is through model stealing attacks, such as model extraction techniques. In this scenario, the adversary only has black-box access to the victim encoder, obtains the embeddings by querying the encoder, and uses the pair of inputs and outputs to learn a piracy encoder with similar functionality and performance to the victim encoder.

\subsubsection{Defender’s Capability and Knowledge} The defender possesses pre-training data as well as the structure and weights of the victim encoder. The defender can use secret methods to protect the victim encoder but cannot modify its parameters. After discovering the suspect encoder, the defender only has black-box access to it without the knowledge of the network structures and datasets used during stealing. Defenders can obtain the embeddings of input images by querying the suspect encoder. Besides, the number of query images should not be too large to avoid being identified as a malicious visitor.

\subsection{Design Overview}
In our work, the StegGuard is proposed as a fingerprint based copyright protection framework for pre-trained encoders, and the overview of StegGuard is shown in Fig. 2. The framework contains three important components: \emph{secrets embeder}, \emph{pre-trained encoder} and \emph{secrets extractor}. The secrets embeder aims to embed the secret into the query image by steganography to obtain the stego image. The pre-trained encoder is employed as the victim encoder and we can access the corresponding embeddings of the input stego images. The secrets extractor is expected to accurately extract the secret from the output embedding. In addition, to reduce the average error of the extracted secrets as well as to improve the visual quality of the stego images, we introduce the Frequency-domain channel attention Embedding block (FcaEmb) into the secrets embeder, which can adaptively select the suitable frequency bands in the image to embed the secret. 

\textbf{\begin{figure}[!ht]
\begin{center}
\includegraphics[width=0.85\linewidth]{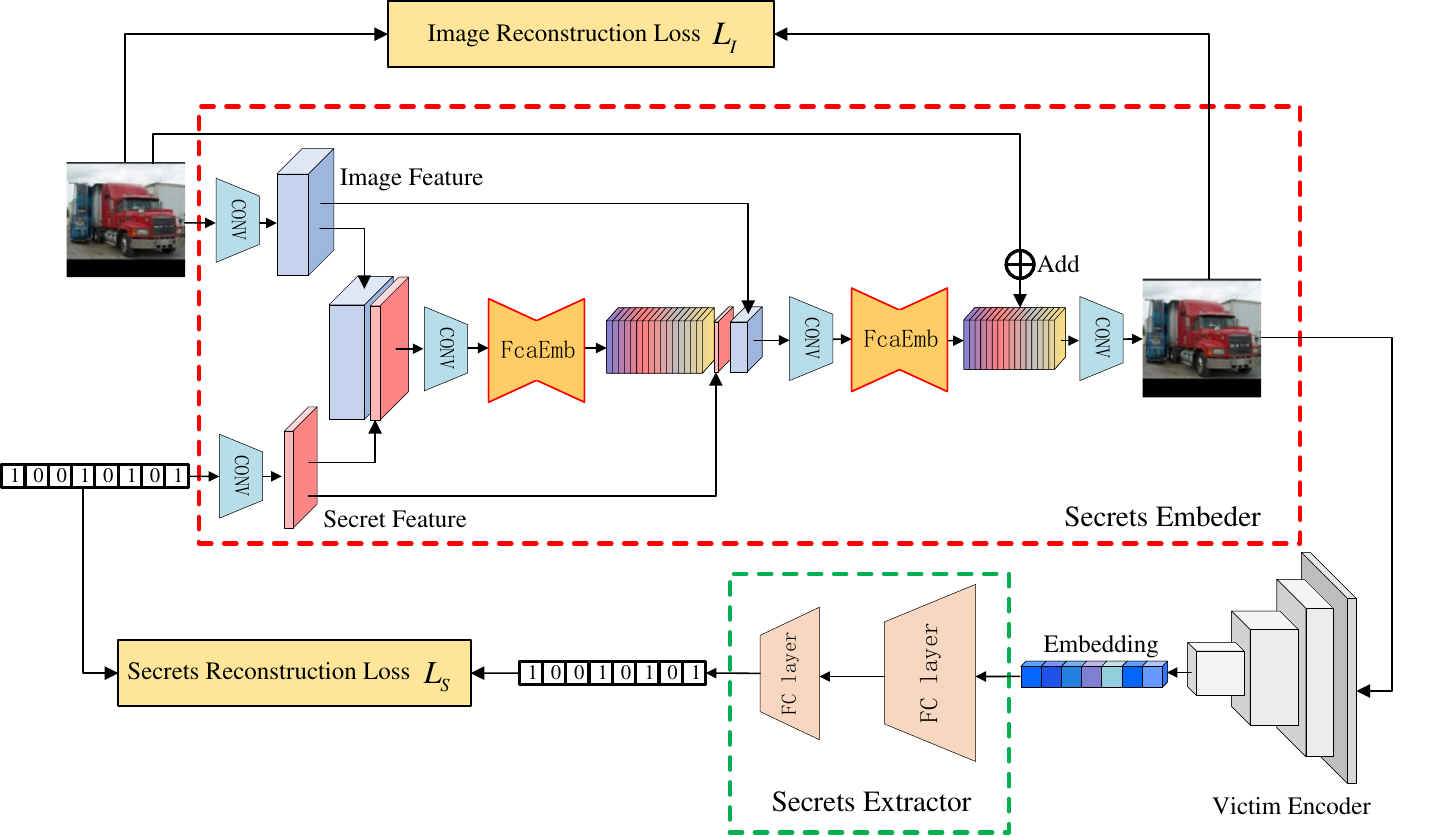}
\caption{The components and pipeline of fingerprint learning involving secrets embeder, secrets extractor and the loss.}
\vspace{-0.7cm}
\end{center}
\end{figure}}

To acquire the fingerprint of the victim encoder, we construct the entire fingerprint learning pipeline as shown in Fig. 3, and freeze the weights of the victim encoder. Then, the secrets embeder and extractor will be jointly optimized by secrets reconstruction loss $\mathcal{L}_S$ and image reconstruction loss $\mathcal{L}_I$. To verify the ownership of a suspect encoder, the victim encoder will be replaced with the suspect encoder. The query images and secrets are fed into the secrets embeder, which generates stego images. These stego images are then used to query the suspect encoder and obtain the embeddings. Compared with embedded secrets, the number of bits in secrets falsely extracted from the embeddings can verify the suspect encoder's ownership.

We provide definitions for all notations used in this work in Table I.

\begin{table}[!ht]
\caption{Definitions of used notations}
\centering 
\resizebox{\columnwidth}{!}{%
\begin{tabular}{c|c}
\hline\hline
\textbf{Notation} & \textbf{Definition} \\ \hline
\textit{$D_{fp}$} & Training dataset in fingerprint learning \\ \hline
\textit{$D_{q}$} & Query dataset in fingerprint verification \\ \hline
\textit{S,E} & Secrets embeder/extractor \\ \hline
\textit{$F_{\lambda_{pt}}$,$F_{v}$,$F_{s}$,$F_{i}$,$F_{p}$} & Pre-trained/victim/suspect/independent/piracy encoder \\ \hline
\textit{$\mathbf{x}$,$\mathbf{x'}$,$\mathbf{y}$} & \begin{tabular}[c]{@{}c@{}}Input image /stego image /output embedding \end{tabular} \\ \hline
\textit{$\lambda_{pt}$,$\theta_{S}$,$\theta_{E}$} & \begin{tabular}[c]{@{}c@{}}Parameters of pre-trained encoder \\ and secret embeder/extractor\end{tabular} \\ \hline
\textit{$\mathbf{k}_{em}$,$\mathbf{k}_{ex}$} & Embedded/Extracted secrets \\ \hline
\textit{L} & Length of secrets \\ \hline
\textit{$ebr$} & Erroneous bit rate of extracted secrets  \\ \hline\hline
\end{tabular}%
}
\end{table}

\subsection{Fingerprint Learning}

\subsubsection{Secrets Embeder}
Utilizing steganography technique, the secrets embeder $S_{\theta}$ embeds the secret into the query image invisibly to get the stego image. Once the generated stego image is sent to the pre-trained encoder, the embedding output from the encoder contains the fingerprint feature of the encoder, which can be used to verify the encoder's ownership. The secrets embeder consists of three block, namely $ConvBNReLU$, $Conv$ and $FcaEmb$ ($Fca$), where $ConvBNReLU$ represents a combined block containing convolutional layer ($Conv$), batch normalization ($BN$) and $ReLU$ activation function.

Let $D_{fp}:=\left \{ \mathbf{x}_i \right \} _{i=1}^N $ denotes the fingerprint training dataset, where $\mathbf{x}_i \sim \mathcal{X} \in \mathbb{R} ^ {C \times H \times W}$ is the clean image, and $\mathbf{k}_{em}\in \left \{ 0,1 \right \}^L $ is the secret of length $L$. It is worth noting that during learning and verification, we randomly generate a secret of length L for each clean image, with no correspondence between the secret and the image. The secrets embeder attempts to embed the secrets into the training dataset to obtain the stego image set $D'_{fp}=\left \{ \mathbf{x'} \mid \mathbf{x'}=S_\theta\left ( \mathbf{x},\mathbf{k}_{em}\right ) , \mathbf{x} \in D_{fp} \right \} $, and $S_{\theta}$ intend to learn a mapping function $S_{\theta}: \mathcal{X}\to \mathcal{X'}$. In detail, when the clean image $\mathbf{x}$ is fed into the secrets embeder, it will be transformed from the 3-dimensional image into a 64-dimensional feature $\mathbf{f}_a$ through the $ConvBNReLU_{3 \to 64}$ block. Then, we concatenate $\left ( Cat \right ) $ the obtained feature $\mathbf{f}_a$ with the secret $\mathbf{k}_{em}$ and compress the dimension from $64+L$ to $64$, and $\mathbf{k}_{em}$ will be embedded into the image features $\mathbf{f}_a$ in the suitable frequency band through $Fca$ block.

\begin{equation}
    \mathbf{f}_a = ConvBNReLU_{3\to 64}\left ( \mathbf{x} \right ) 
\end{equation}
\begin{equation}
        \mathbf{f}_b = Fca\left ( ConvBNReLU_{64+L\to 64}\left ( Cat\left (\mathbf{f}_a, \mathbf{k}_{em} \right )  \right ) \right ) 
\end{equation}

Next, $\mathbf{f}_a$, $\mathbf{f}_b$ and $\mathbf{k}_{em}$ are concatenated in the same way, and $\mathbf{f}_c$ will be obtained after $ConvBNReLU_{64+64+L\to 64}$ and $Fca$ module. Finally, in order to get a stego image $\mathbf{x'}$ with the same size as the original image $\mathbf{x}$, we need to transform the concatenation of $\mathbf{f}_c$ and $\mathbf{x}$ to $\mathbb{R} ^ {3 \times H \times W}$ through the $Conv$ block and the formula is shown as follows: 

\begin{equation}
    \mathbf{f}_c=Fca\left ( ConvBNReLU_{64+64+L\to 64}\left ( Cat\left ( \mathbf{f}_b, \mathbf{k}_{em} ,\mathbf{f}_a\right )  \right ) \right ) 
\end{equation}
\begin{equation}
    \mathbf{x'}=S_{\theta}\left ( \mathbf{x} \right ) =Conv_{64+3\to 3}\left ( Cat\left ( \mathbf{f}_c, \mathbf{x}\right )  \right )
\end{equation}

\subsubsection{Secrets Extractor}
After sending the stego image $\mathbf{x'}$ into the pre-trained encoder $F_{\lambda_{pt}}$, the embedding $\mathbf y = F_{\lambda_{pt}}\left( \mathbf{x'} \right ),\, \mathbf{y}\in \mathbb{R} ^ D$ containing rich feature of the encoder can be obtained, where $D$ is the dimension of the embedding. The function of secrets extractor $E_{\theta}$ is to compress the complex features and transform them into an understandable secret. In order to improve the representational capability of the extractor, two fully connected layers $\left (FC  \right ) $ and $ReLU$ function are used as the structure of the extractor. To extract the secret from the embedding, we feed the embedding $\mathbf{y}$ into the secrets extractor and after the transformation of $\left (FC  \right ) $ layer and $ReLU$ function, we can finally get the extracted secret $\mathbf{k}_{ex}$ with the same length as the embedded secret $\mathbf{k}_{em}$  and the formula is shown as follows: 

\begin{equation}
\begin{aligned}
\mathbf{k}_{ex} = E_{\theta}\left ( \mathbf{y} \right ) =FC\_ReLU_{256\to L}\left ( FC\_ReLU_{D\to 256} \left ( \mathbf{y} \right )  \right ) 
\end{aligned}
\end{equation}

\subsubsection{Loss Function}

In StegGuard, the secrets embeder and extractor are jointly optimized, which is controlled in two loss functions: secrets reconstruction loss $\mathcal{L}_S$ and image reconstruction loss $\mathcal{L}_I$.

We propose the secrets reconstruction loss function $\mathcal{L}_S$ for the secrets embeder and extractor to embed and extract the secrets correctly. It ensures that each bit in the secret $\mathbf{k}_{ex}$ extracted from embedding and the secret $\mathbf{k}_{em}$ embedded in stego image are aligned by reducing the difference between them. Besides, it is also necessary to ensure that the stego image is still visually similar to the original image to prevent the adversary from filtering the query images and rejecting anomalous queries. Thus, we adopt image reconstruction loss $\mathcal{L}_I$, which reduces the difference between the original image $\mathbf{x}$ and the stego image $\mathbf{x'}$. The definitions of the loss function are as follows:

\begin{equation}
\mathcal{L}_S =\mathbb{E}_{S}\frac{1}{L} \left \| \mathbf{k}_{em}- \mathbf{k}_{ex} \right \| _{2}^{2} 
\end{equation}

\begin{equation}
    \mathcal{L}_I = \mathbb{E}_{I}\frac{1}{3\times W\times H} \left \| \mathbf{x}- \mathbf{x'} \right \| _{2}^{2} 
\end{equation}

 $\mathcal{L}_S$ and $\mathcal{L}_I$ is balanced by the weight factor $\alpha$ and we use 0.7 as default. Therefore, the secrets embeder and extractor are jointly optimized by minimizing the following loss function:
\begin{equation}
\underset{\theta_{S},\theta_{E}}{\mathrm{minimize}} \ \mathcal{L}_I+  \alpha \cdot \mathcal{L}_S 
\end{equation}

\textbf{\begin{figure}[!ht]
\begin{center}
\includegraphics[width=0.9\linewidth]{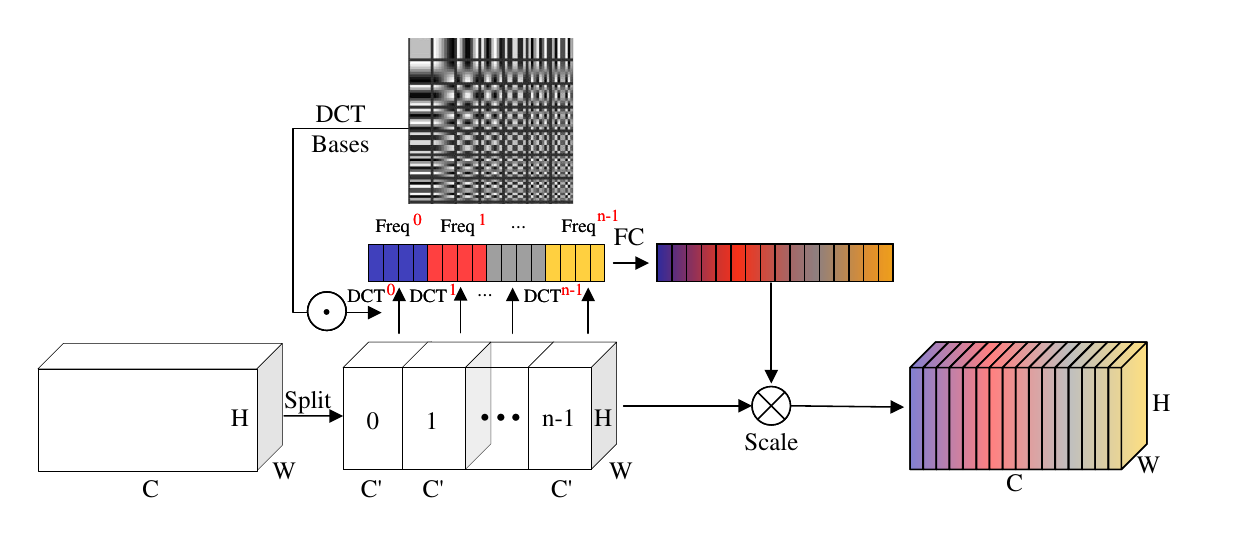}
\caption{The structure of FcaEmb block.}
\end{center}
\end{figure}}

\subsubsection{FcaEmb}
Existing image steganography methods tend to embed the message with fewer parameters and simpler network structures to speed up the convergence of the model, but these strategies lead to a concentration of the message embedded in the low-frequency band, while the middle and high-frequency bands are not fully utilized. Since the middle and high-frequency bands of the query dataset will also be used as important features in verifying the ownership of the encoder. Therefore, in our method, we introduce the Frequency-domain Channel Attention embedding block (FcaEmb) \cite{paper36}, which can fully utilize the features of each frequency band. While improving the embedding efficiency, it enables the copyright owner to learn the encoder's fingerprint more accurately. The structure of the module is shown in Fig. 4.

When the input feature tensor $ \left \{ {\mathbf{x}^j}\right \}_{j=1}^C  $ has $C$ channels, we slice each channel to get the channel-wise of the tensor $\mathbf{x} ^ {i} \in \mathbb{R} ^ {H\times W}$, $\mathit{i} \in \left \{ 0,1,\cdots ,C-1 \right \} $. Then, each slice will be mapped into the 2D-DCT domain, and for computational efficiency, each dimension will be divided into 7 equal parts which have a size of $7\times 7$. Specifically, we obtain the 49-dimensional vector $\mathit{DCT}^{7 \times 7} \in \mathbb{R} ^ {49}$ , the basis function of 2D-DCT is shown as follows:

\begin{equation}
    B_{h, w}^{i, j}=\cos \left(\frac{\pi h}{H}\left(i+\frac{1}{2}\right)\right) \cos \left(\frac{\pi w}{W}\left(j+\frac{1}{2}\right)\right)
\end{equation}

\begin{equation}
\begin{split}
    Freq^{i}=2 D D C T^{u, v}\left(\mathbf{x}^{i}\right), u=i / 7, v=i\ mod \ 7 \\
    =\sum_{h=0}^{H-1} \sum_{w=0}^{W-1} \mathbf{x}_{:, h, w}^{i} B_{h, w}^{u,v} \text { s.t. } i \in\{0,1, \cdots, n-1\} 
\end{split}
\end{equation}

\begin{equation}
    Freq^{7 \times 7}=Cat\left(\left[Freq^{0}, Freq^{1}, \cdots, Freq^{48}\right]\right)
\end{equation}

Then, the 49-dimension vector $\mathit{DCT}^{7 \times 7} \in \mathbb{R} ^ {49}$ needs to be dimensionalized by the fully connected layer $\mathit{FC}$, while the weight vector $\mathbf{v}$ is obtained by $\mathbf{v}=sigmoid\left ( FC\left ( Freq \right )  \right ) $.

Finally, the weights $\mathbf{v}$ and feature tensor $\mathbf{x}$ will be multiplied by the scale function; in this way, we can make full use of the frequency bands of the query image and improve the visual invisibility of the stego image. And the enhanced tensor $\mathbf{\widetilde{x}}=F_{scale}\left ( \mathbf{x},\mathbf{v} \right ) $.

\emph{\textbf{Remark:}} In this work, we use forward optimization to make the output result $\mathbf{k}_{ex}$ be consistent with $\mathbf{k}_{em}$. In fact, we can also achieve the effect of copyright verification by optimizing the loss function in opposite direction, so that $\mathbf{k}_{ex}$ and $\mathbf{k}_{em}$ will be completely reversed, which maximizes the $ebr$. To unify the secrets reconstruction loss and image reconstruction loss, we adopt the forward optimization for secrets reconstruction loss.

\begin{algorithm}
\caption{Fingerprint learning and verification}
\label{algorithm:Fingerprint}
\SetKwData{In}{\textbf{in}}\SetKwData{To}{to}
\DontPrintSemicolon
\SetAlgoLined
\KwIn {pre-trained victim encoder $F_{\lambda_{pt}}$, suspect encoder $F_{s}$, fingerprint learning dataset $D_{fp}$, query dataset $D_q$, embedded secret $\mathbf{k}_{em}$, threshold $T$, learning rate $\gamma$.}
\KwOut {secrets embeder $S$, secrets extractor $E$, extracted secret $\mathbf{k}_{ex}$.}
  /* $\qquad$ Fingerprint learning $\qquad$ */ \;
  \While{ $ebr \ge T$ }
  {
   $ebr$ = 0 \;
  \For{$\mathbf{X}_i$ in $D_{fp}=\left \{ \mathbf{X}_1,\mathbf{X}_2,\dots,\mathbf{X}_M  \right \}$}{
      /*$\mathbf{X}_i$ denotes a batch of images*/ \;
      Randomly generate a batch of secrets $\mathbf{K}_{em}^i$\;
      
      $\mathbf{X'}_i$ = $S_\theta\left (\mathbf{X}_i, \mathbf{K}_{em}^i \right )$ 
      $\triangleright $ obtain stego images\;
      
      $\mathbf{Y} = F_{\lambda_{pt}}\left ( \mathbf{X'}_i  \right )$ 
      $\qquad\triangleright $ obtain embeddings for stego images\;
      
      $\mathbf{K}_{ex}^i = E_\theta\left (\mathbf{Y} \right ) $ 
      $\qquad\,\,  \triangleright $ extract secrets \;

      /*Equation (6),(7) provide the formulas for $\mathcal {L}_{S}$ and $\mathcal {L}_{I}$*/ \;
      $\mathcal{L}_{total}$ = $ \mathcal {L}_{S}^i  \left ( \mathbf{K}_{em}^i, \mathbf{K}_{ex}^i \right ) + \alpha \cdot \mathcal{L}_{I}^i  \left (\mathbf{X}_i, \mathbf{X'}_i \right )$ \;
      
      $\theta _{i+1}=\theta_{i}- \gamma \cdot \bigtriangledown _{\theta}\mathcal{L}_{total} $ $\quad\triangleright $update the parameters\;
      
      $ebr$ = $ebr$ + $\mathcal {L}_{S}^i  \left ( \mathbf{K}_{em}^i, \mathbf{K}_{ex}^i \right )$ \;
  }
  
  $ebr$ = $ebr$ / $M$ \;
  }

  /* $\qquad$ Fingerprint verification $\qquad$ */ \;
  $ebr$= 0 \;
  \For{$\mathbf{x}_j$ in $D_q=\left \{ \mathbf{x}_1,\mathbf{x}_2,\dots,\mathbf{x}_N  \right \} \in D_{fp}, N  \ll \left | D_{fp} \right | $}{
       Randomly generate a secret $\mathbf{k}_{em}^j$\;
      $\mathbf{k}_{ex}^j = E_\theta\left ( F_s\left ( S_\theta\left (  \mathbf{x}_j, \mathbf{k}_{em}^j\right )  \right )  \right ) $\;
      $ebr$ = $ebr$ + $\left \| \mathbf{k}_{em}^j - \mathbf{k}_{ex}^j \right \| ^2_2$ \;
  }
  $ebr$ = $ebr$ / $N$ \;

  \eIf{$ebr < T$}
  {The suspect encoder is the piracy encoder}
  {The suspect encoder is the independent encoder}
\end{algorithm}

\subsection{Fingerprint Verification}

When the encoder owner obtains the corresponding fingerprint, it can be used to verify the ownership of the suspect encoder. First, we only need to select a very small number of images from the fingerprint training dataset as the query dataset $D_q \subset  D_{fp} \left ( \left | D_q \right | \ll \left | D_{fp} \right |  \right ) $. Then the secrets will be embedded into these query images through the secrets embeder to acquire the corresponding stego query set $D'_q := \left \{ \mathbf{x'}_{i} \right \}^M_{i=1} $. Second, $D'_q$ will be sent into the suspect encoder, and we can obtain the embeddings of the stego images. Finally, the secrets will be further extracted from embeddings by the secrets extractor and compared with the embedded secrets to calculate $ebr$. The ownership of the suspect encoder can be verified by comparing $ebr$ with threshold $T$, whether the suspect encoder is piracy or independent, which is analyzed as follows:

If $ebr$ is less than the threshold $T$, the suspect encoder can be verified as a piracy encoder. In other words, a lower $ebr$ indicates that the suspect encoder is more similar to the victim encoder. If the encoder is an independent encoder, the extracted secrets will be randomly guessed, so $ebr$ is around 0.5. The complete learning and verification process is shown in Algorithm 1.

\section{Experiment and Analysis}

\subsection{Experimental Setup}

\textbf{Datasets}. We use a total of four datasets including ImageNet \cite{paper37}, CIFAR-10 \cite{paper38}, GTSRB \cite{paper39}, and STL-10 \cite{paper40}. Specially, we randomly select 50 semantic categories from the ImageNet dataset, and we also use STL-10 to train the secrets embeder and extractor, as well as to verify the ownership of the suspect encoder.

\textbf{Pre-trained Encoder}. To verify the performance of StegGuard, we select different SSL algorithms and model structures to perform extensive evaluations. We use three self-supervised learning algorithms including SimCLR, MoCo v2 and MAE to pre-train the encoder and use ResNet-18 \cite{paper41}, ResNet-34, ResNet-50, VGG-16 \cite{paper42}, DenseNet-121 \cite{paper43}, ViT-B \cite{paper44}, ViT-L and ViT-H networks as model structures of pre-trained encoder.

Meanwhile, the total number of epochs for the pre-training SSL encoder is set to 200, and the number of epochs for learning the secrets embeder and extractor is set to 100 by default. Additionally, we conduct simulations with PyTorch, accelerated by a single RTX 3090 GPU.

\subsection{Effectiveness}

Given learned fingerprint and query set, the purpose of this experiment is to evaluate the distinguishment between victim encoder and independent encoder. We get three victim encoders differing from each other owing to different model structure, different pre-training dataset and different SSL algorithm(See Table II, individual victim is denoted as A,B,C, respectively). In addition, we design two types of independent encoders that differ from victim encoder along two dimensions. One dimension is the same as the difference between individual tested victim encoder(See section IV B.1); the other dimension is hyper-parameter variation(See section IV B.2). As concerns the setting of fingerprint and query set, it is listed in Table III. 

\begin{table}[!ht]
    \centering
    \caption{The training setting of the victim encoders}
    \resizebox{0.86\columnwidth}{!}{%
    \begin{tabular}{c  c  c  c }
    \hline\hline
        \multicolumn{1}{c|}{Encoder} & \multicolumn{1}{c|}{Model Structure} & \multicolumn{1}{c|}{ Pre-training Dataset } & SSL Algorithm  \\ \hline
        \multicolumn{1}{c|}{Victim A} & \multicolumn{1}{c|}{ ResNet-50 } & \multicolumn{1}{c|}{ ImageNet} &  SimCLR \\ 
        \multicolumn{1}{c|}{Victim B} & \multicolumn{1}{c|}{ ResNet-18 }& \multicolumn{1}{c|}{CIFAR-10} & MoCo v2\\ 
        \multicolumn{1}{c|}{Victim C} &\multicolumn{1}{c|}{ ViT-B} &\multicolumn{1}{c|}{ ImageNet} & MAE \\ \hline\hline
    \end{tabular}%
    }
\end{table}
\begin{table}[!ht]
\centering
\caption{The setting of fingerprint and query set}
\resizebox{\columnwidth}{!}{
\begin{tabular}{l|c|c|c|c}
\hline\hline
\multirow{2}{*}{Fingerprint} & Dataset & \#Training Data & \#Query Data & Secrets Length \\ \cline{2-5} 
 & STL-10 & Full & 1K & 64 \\ \hline\hline
\end{tabular}}
\end{table}

\subsubsection{Detection Independent Encoders Varying with Model Structure, Pre-training Dataset and SSL Algorithm }

 Relying on the three factors involving model structure, pre-training dataset and SSL algorithm, we get multiple independent encoders differing from victim encoder by altering one factor and keeping the other two factors preserved. As shown in Fig. 5, $ebr$ is much lower than 0.5 for all the evaluated victims. In contrast, $ebr$ is around 0.5 for all the evaluated independent encoders and much higher than that of the victim encoders. Some results are missing because that data is unavailable, for example, the model structure of victim C is ViT-B so that only MAE algorithm could be used(See Fig. 5(c)). 

\textbf{\begin{figure*}[!ht]
\centering
\includegraphics[width=0.76\linewidth]{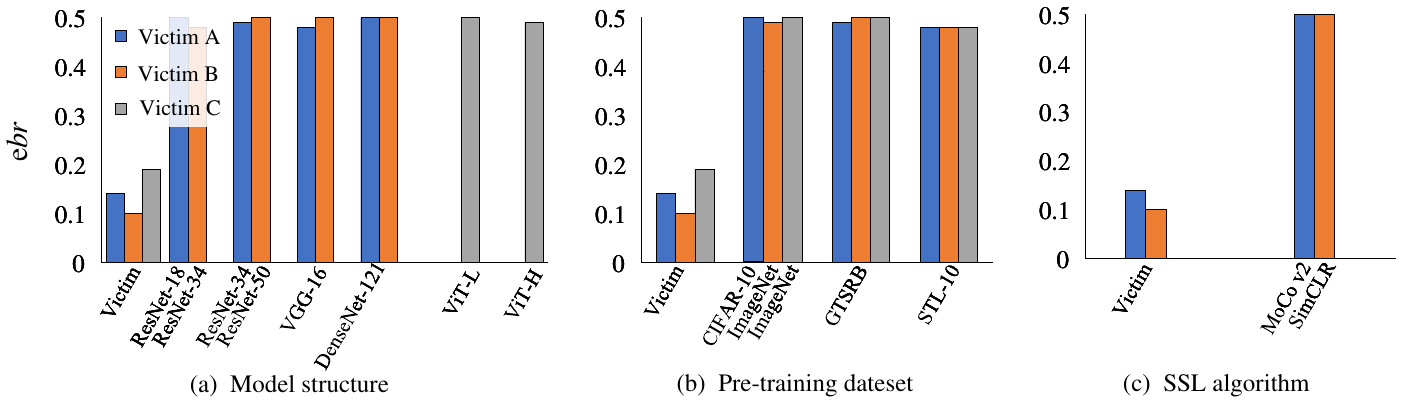}
\caption{The effectiveness of detecting independent encoders varying with model structure, pre-training dataset and SSL algorithm.}
\end{figure*}}
\textbf{\begin{figure*}[!ht]
\centering
\includegraphics[width=0.78\linewidth]{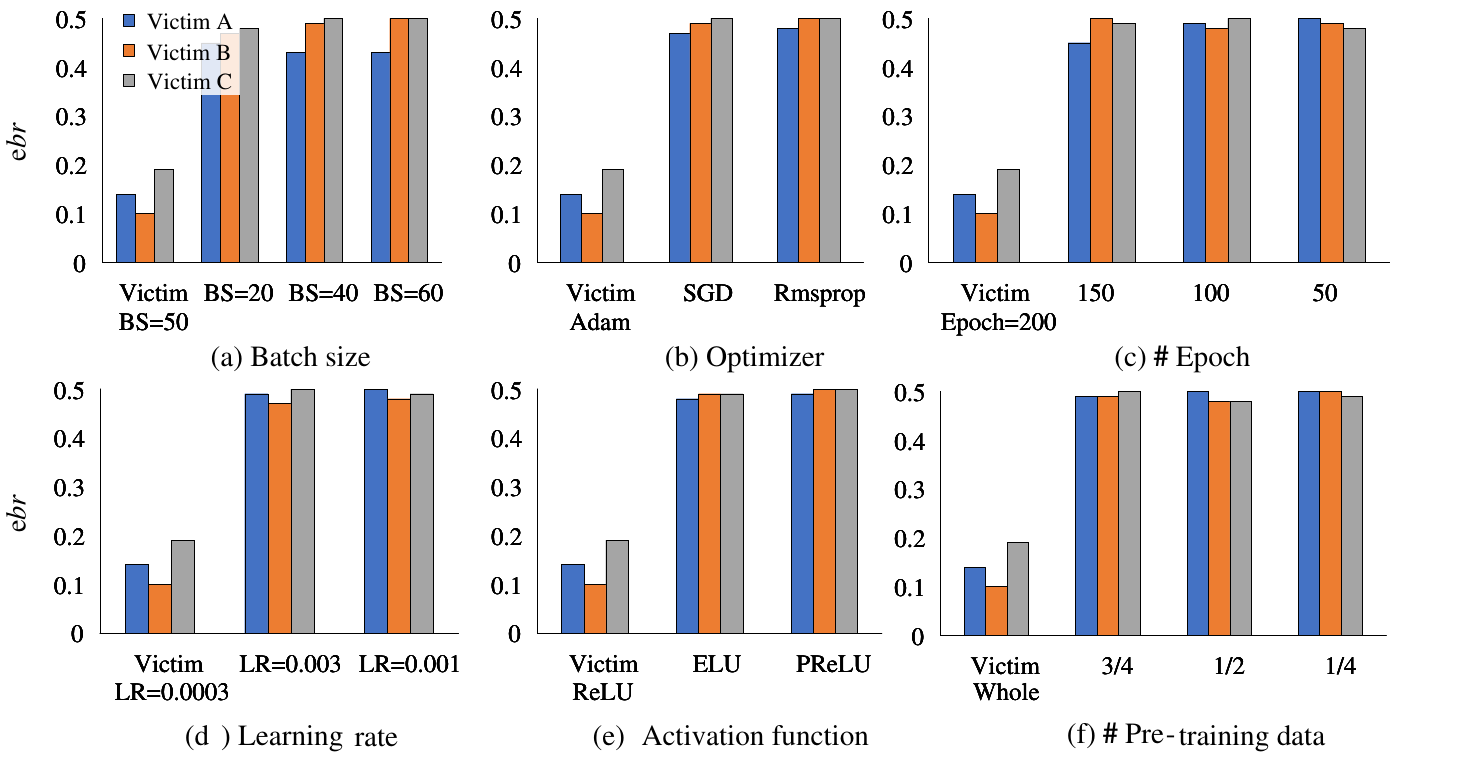}
\caption{The effectiveness of detecting independent encoders varying with hyper-parameters.}
\end{figure*}}

\subsubsection{Detection Independent Encoders Varying with Hyper-parameters}

Besides the major difference discussed above, independent encoder also can be different from victim encoder by merely changing one hyper-parameter while the other hyper-parameters remain unchanged. The tested hyper-parameters include batch size, optimizer, number of epoch, learning rate, activation function, and number pre-training data. The experimental results in Fig. 6(a-f) demonstrate that $ebr$ is rather sensitive to the alteration of hyper-parameter and $ebr$ of each hyper-parameter altered independent encoder will rise around 0.5.

\subsection{Robustness}

Given learned fingerprint and query set(Detailed list in Table III), the purpose of this experiment is to evaluate the robustness of piracy encoder detection. Through model extraction, we get a piracy encoder with comparable performance to the victim encoder. Additionally, we simulate infringement behavior by fine-tuning or pruning victim encoder, noising and shuffle embedding of victim encoder. 

\subsubsection{Model Extraction}

Adversaries sometimes cannot obtain the victim encoders deployed on the cloud platform directly, and they will adopt model extraction technique, a model stealing attack, to get the encoding performance of the victim encoders. Adversaries usually query the victim encoders through different surrogate datasets and model structures. The pair of the input images and the output embeddings will be used to train a piracy encoder which has similar functions and performance to the victim encoder, which makes it difficult to verify the ownership of the piracy encoder. 

To evaluate the robustness of the StegGuard against model extraction attack, we use different model structures and datasets to steal the victim encoder. The experimental results are shown in Fig. 7 and each dimension in the radar plot represents the different structure of the surrogate models used to steal the victim encoders, including ResNet-18, ResNet-34, ResNet-50, VGG-16 and DenseNet-121, respectively. In addition, we use the solid line to represent $ebr$ of the piracy encoder, while the dotted line represents the victim encoder. The experimental results show that StegGuard is effective in verifying piracy encoders extracted through different datasets and model structures. For comparison, we also select other independent encoders under model extraction attack and the experimental results are shown in the dashed line in the Fig. 7, which indicates that the secrets cannot be extracted correctly from these encoders.

\textbf{\begin{figure}[!t]
\begin{center}
\includegraphics[width=\linewidth]{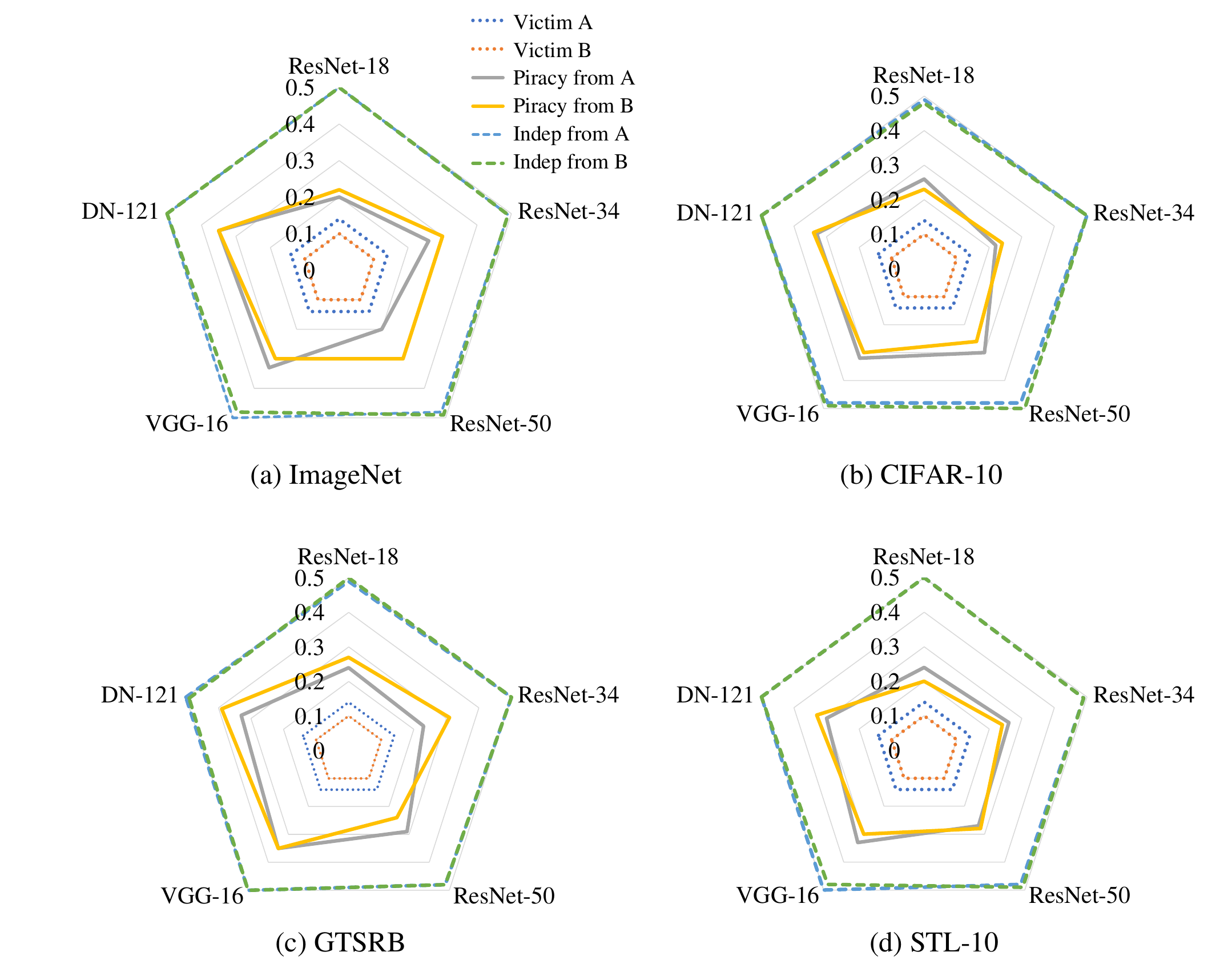}
\caption{Robustness of the StegGuard against model extraction attack. (a$\sim $d) show different surrogate datasets and structures used to steal the victim encoders.}
\vspace{-0.8cm}
\end{center}
\end{figure}}

In order to comprehensively evaluate the detection performance of this method for piracy encoders. We randomly select 40 encoders from a large number of pre-trained encoders, including 20 piracy encoders and 20 independent encoders. To precisely evaluate the performance of StegGuard, we adopt two evaluation metrics, denoted by $\triangle \mu$ and p-value$\left ( \in \left [ 0,1 \right ] \right ) $, where $\triangle \mu$ denotes the difference between the piracy encoders and the independent encoders in log-likelihoods. Further more, we adopt p-values to evaluate the performance of the fingerprints. Specifically, we set $\Omega $ and $\Omega_{Indep}$ be two sets of the secrets from the suspect encoders and independent encoders and the null hypothesis as: $\mathcal{H}_0:\mu < \mu_{Indep}$, where $\mu=\bar{\Omega },\mu_{Indep}=\bar{\Omega}_{Indep} $. We set the predefined significance level $\alpha$=0.05 and the $t$-Test can reject $\mathcal{H}_0$ to claim the piracy encoder. In addition, a lower value of the p-value indicates that the encoder is more similar to the victim encoder, on the contrary, the independent encoders are high. Conversely, the piracy encoders will have a large $\triangle \mu$ than the independent encoders. 

To compare with DI\cite{paper29}, we employ the same model extraction attack technique \cite{paper34} and perform a detailed comparison under three victim encoders and three datasets. The experimental results are shown in Table IV, the $F_v$, $F_p$ and $F_i$ denote the victim encoder, piracy encoder and independent encoder, respectively. And `-' in the table means the untestable scenarios. The empirical results in Table IV demonstrate that StegGuard has a lower p-value and a higher $\triangle \mu$ when verifying piracy encoder and the independent encoder is the opposite.

\begin{table}[!ht]
\centering
\caption{Comparison of StegGuard and DI in terms of the p-values and $\triangle \mu$}
\resizebox{\columnwidth}{!}{%
\begin{tabular}{ccc|cc|cc|cc}
\hline\hline
\multicolumn{3}{c|}{} & \multicolumn{2}{c|}{CIFAR-10} & \multicolumn{2}{c|}{ImageNet} & \multicolumn{2}{c}{STL-10} \\ \cline{4-9} 
\multicolumn{3}{c|}{\multirow{-2}{*}{Victim Encoder}} & \multicolumn{1}{c|}{p-value} & $\triangle \mu$ & \multicolumn{1}{c|}{p-value} & $\triangle \mu$ & \multicolumn{1}{c|}{p-value} & $\triangle \mu$ \\ \hline
\multicolumn{1}{c|}{} & \multicolumn{1}{c|}{} & $F_v$ & \multicolumn{1}{c|}{\cellcolor[HTML]{C0C0C0}-} & \cellcolor[HTML]{C0C0C0}- & \multicolumn{1}{c|}{\cellcolor[HTML]{C0C0C0} 0} & \cellcolor[HTML]{C0C0C0}16.32 & \multicolumn{1}{c|}{\cellcolor[HTML]{C0C0C0}-} & \cellcolor[HTML]{C0C0C0}- \\
\multicolumn{1}{c|}{} & \multicolumn{1}{c|}{} & $F_p$ & \multicolumn{1}{c|}{$10^{-11}$} & 10.61 & \multicolumn{1}{c|}{$10^{-14}$} & 12.35 & \multicolumn{1}{c|}{$10^{-10}$} & 10.10 \\
\multicolumn{1}{c|}{} & \multicolumn{1}{c|}{\multirow{-3}{*}{StegGuard}} & $F_i$ & \multicolumn{1}{c|}{0.96} & 0.13 & \multicolumn{1}{c|}{0.90} & 0.18 & \multicolumn{1}{c|}{0.95} & 0.15 \\ \cline{2-9} 
\multicolumn{1}{c|}{} & \multicolumn{1}{c|}{} & $F_v$ & \multicolumn{1}{c|}{\cellcolor[HTML]{C0C0C0}-} & \cellcolor[HTML]{C0C0C0}- & \multicolumn{1}{c|}{\cellcolor[HTML]{C0C0C0}0} & \cellcolor[HTML]{C0C0C0}8.02 & \multicolumn{1}{c|}{\cellcolor[HTML]{C0C0C0}-} & \cellcolor[HTML]{C0C0C0}- \\
\multicolumn{1}{c|}{} & \multicolumn{1}{c|}{} & $F_p$ & \multicolumn{1}{c|}{$10^{-4}$} & 4.38 & \multicolumn{1}{c|}{$10^{-7}$} & 6.91 & \multicolumn{1}{c|}{$10^{-6}$} & 4.62 \\
\multicolumn{1}{c|}{\multirow{-6}{*}{\begin{tabular}[c]{@{}c@{}} Victim A\end{tabular}}} & \multicolumn{1}{c|}{\multirow{-3}{*}{DI\cite{paper29}}} & $F_i$ & \multicolumn{1}{c|}{0.68} & 0.40 & \multicolumn{1}{c|}{-} & - & \multicolumn{1}{c|}{0.72} & 0.36 \\ \hline
\multicolumn{1}{c|}{} & \multicolumn{1}{c|}{} & $F_v$ & \multicolumn{1}{c|}{\cellcolor[HTML]{C0C0C0}0} & \cellcolor[HTML]{C0C0C0}25.93 & \multicolumn{1}{c|}{\cellcolor[HTML]{C0C0C0}-} & \cellcolor[HTML]{C0C0C0}- & \multicolumn{1}{c|}{\cellcolor[HTML]{C0C0C0}-} & \cellcolor[HTML]{C0C0C0}- \\
\multicolumn{1}{c|}{} & \multicolumn{1}{c|}{} & $F_p$ & \multicolumn{1}{c|}{$10^{-12}$} & 18.30 & \multicolumn{1}{c|}{$10^{-10}$} & 14.22 & \multicolumn{1}{c|}{$10^{-10}$} & 15.16 \\
\multicolumn{1}{c|}{} & \multicolumn{1}{c|}{\multirow{-3}{*}{StegGuard}} & $F_i$ & \multicolumn{1}{c|}{0.99} & 0.03 & \multicolumn{1}{c|}{0.93} & 0.12 & \multicolumn{1}{c|}{0.89} & 0.19 \\ \cline{2-9} 
\multicolumn{1}{c|}{} & \multicolumn{1}{c|}{} & $F_v$ & \multicolumn{1}{c|}{\cellcolor[HTML]{C0C0C0}0} & \cellcolor[HTML]{C0C0C0}20.42 & \multicolumn{1}{c|}{\cellcolor[HTML]{C0C0C0}-} & \cellcolor[HTML]{C0C0C0}- & \multicolumn{1}{c|}{\cellcolor[HTML]{C0C0C0}-} & \cellcolor[HTML]{C0C0C0}- \\
\multicolumn{1}{c|}{} & \multicolumn{1}{c|}{} & $F_p$ & \multicolumn{1}{c|}{$10^{-7}$} & 7.25 & \multicolumn{1}{c|}{$10^{-7}$} & 7.94 & \multicolumn{1}{c|}{$10^{-5}$} & 5.13 \\
\multicolumn{1}{c|}{\multirow{-6}{*}{\begin{tabular}[c]{@{}c@{}} Victim B \end{tabular}}} & \multicolumn{1}{c|}{\multirow{-3}{*}{DI}} & $F_i$ & \multicolumn{1}{c|}{-} & - & \multicolumn{1}{c|}{0.90} & 0.26 & \multicolumn{1}{c|}{0.81} & 0.35 \\ \hline
\multicolumn{1}{c|}{} & \multicolumn{1}{c|}{} & $F_v$ & \multicolumn{1}{c|}{\cellcolor[HTML]{C0C0C0}-} & \cellcolor[HTML]{C0C0C0}- & \multicolumn{1}{c|}{\cellcolor[HTML]{C0C0C0}0} & \cellcolor[HTML]{C0C0C0}22.12 & \multicolumn{1}{c|}{\cellcolor[HTML]{C0C0C0}-} & \cellcolor[HTML]{C0C0C0}- \\
\multicolumn{1}{c|}{} & \multicolumn{1}{c|}{} & $F_p$ & \multicolumn{1}{c|}{$10^{-8}$} & 9.15 & \multicolumn{1}{c|}{$10^{-10}$} & 12.54 & \multicolumn{1}{c|}{$10^{-8}$} & 8.80 \\
\multicolumn{1}{c|}{} & \multicolumn{1}{c|}{\multirow{-3}{*}{StegGuard}} & $F_i$ & \multicolumn{1}{c|}{0.95} & 0.10 & \multicolumn{1}{c|}{0.93} & 0.13 & \multicolumn{1}{c|}{0.94} & 0.11 \\ \cline{2-9} 
\multicolumn{1}{c|}{} & \multicolumn{1}{c|}{} & $F_v$ & \multicolumn{1}{c|}{\cellcolor[HTML]{C0C0C0}-} & \cellcolor[HTML]{C0C0C0}- & \multicolumn{1}{c|}{\cellcolor[HTML]{C0C0C0}0} & \cellcolor[HTML]{C0C0C0}14.37 & \multicolumn{1}{c|}{\cellcolor[HTML]{C0C0C0}-} & \cellcolor[HTML]{C0C0C0}- \\
\multicolumn{1}{c|}{} & \multicolumn{1}{c|}{} & $F_p$ & \multicolumn{1}{c|}{$10^{-4}$} & 4.23 & \multicolumn{1}{c|}{$10^{-5}$} & 5.18 & \multicolumn{1}{c|}{$10^{-5}$} & 4.87 \\
\multicolumn{1}{c|}{\multirow{-6}{*}{\begin{tabular}[c]{@{}c@{}} Victim C \end{tabular}}} & \multicolumn{1}{c|}{\multirow{-3}{*}{DI}} & $F_i$ & \multicolumn{1}{c|}{0.51} & 0.69 & \multicolumn{1}{c|}{-} & - & \multicolumn{1}{c|}{0.60} & 0.55 \\ \hline\hline
\end{tabular}%
}
\end{table}

\subsubsection{Fine-tuning}

The adversaries often adopt some model modification attacks to avoid the stolen encoders from being detected. Therefore, we fine-tune the SSL encoder in two ways, FT-Same and FT-Other, where FT-Same means fine-tuning all parameters of the encoder with the same dataset as the pre-training dataset, and FT-Other means fine-tuning with another dataset different from the pre-training dataset. `FT-Other' in Fig. 8 means that the victim A,B,C are fine-tuned by GTSRB dataset. The experimental results are shown in Fig. 8, which indicates that as the epoch number of encoder's fine-tuning increases, $ebr$ of the extracted secrets also rises, and the fingerprint is more vulnerable to FT-Other attack. When the number of fine-tuning epochs increases to about 80, $ebr$ tends to stabilize and is still effective in verifying the piracy encoders.

\textbf{\begin{figure}[!ht]
\begin{center}
\includegraphics[width=0.7\linewidth]{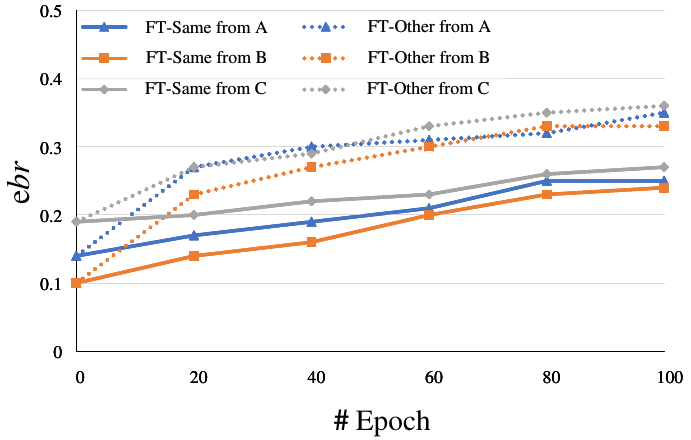}
\caption{Robustness of the StegGuard against different fine-tuning methods.}
\end{center}
\end{figure}}

\subsubsection{Pruning}

Pruning is an effective model compression technique that is also widely used to destroy the fingerprint of the model. By pruning some neurons, it makes the encoder unable to extract the correct embeddings from the stego images. In our experiments, we use the smallest L1-norms to remove the neurons in the encoder. The experimental results are shown in Fig. 9. When the pruning rate is below 0.4, StegGuard can maintain a low $ebr$ in verifying the piracy encoder. And when the pruning rate is 0.4 and above, $ebr$ will increase, but it is still able to verify the encoder's ownership effectively.

\textbf{
\begin{figure}[!ht]
    \begin{minipage}[t]{0.49\linewidth}
    \centering
    \includegraphics[scale=0.36]{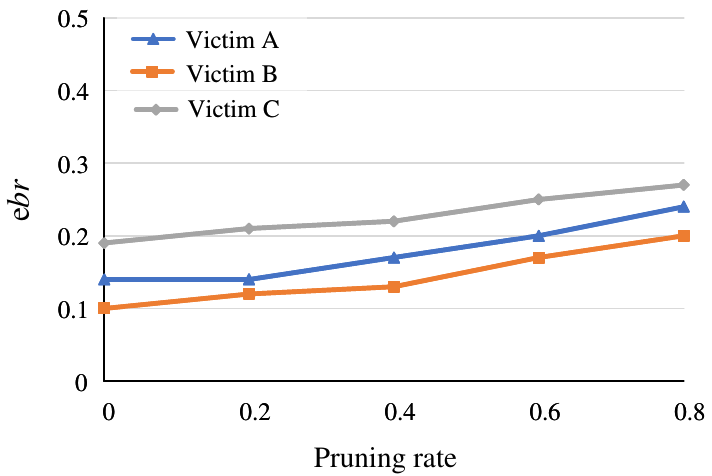}
    \caption{Robustness of the StegoGuard against pruning}
    \end{minipage}
    \begin{minipage}[t]{0.49\linewidth}
    \centering
    \includegraphics[scale=0.37]{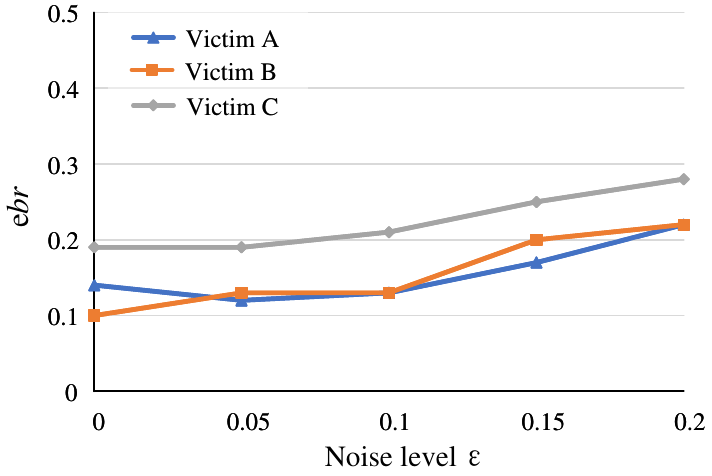}
    \caption{Robustness of the StegoGuard against embedding noising}
    \end{minipage}
\end{figure}
}

\subsubsection{Embedding Noising and Shuffle}

To hinder the copyright owner from extracting secrets correctly from the output embeddings. The adversary can add Gaussian noise to the output embedding as $h'=h+  \varepsilon \cdot \mathcal{N} \left ( 0,1 \right ) $,  where $h$ is the original embedding, $h'$ is the perturbed embedding, and $\varepsilon$ is a hyper-parameter to control the noise level. The experimental results are shown in Fig. 10. When the strength of $\varepsilon $ increases to 0.15, it will have some impact on the extraction of the secrets, but the defender can still efficiently verify the ownership of the encoder. If the adversary further enhances the strength of the noise, this will cause more damage to the embeddings and weaken the performance of the encoder. Moreover, we use $t$-Test to evaluate the robustness of StegGuard and the results are shown in Table V. As a complement to the noising robustness, we also tested the robustness to embedding shuffle. Specifically, we randomly shuffle 5\% of each embedding and feed them into the secrets extractor. Experimental results show that StegGuard can still effectively extract secrets and verify piracy encoders.

\begin{table}[!ht]
\caption{The robustness of StegGuard}
\centering
\resizebox{\columnwidth}{!}{%
\begin{tabular}{c|cccccc}
\hline\hline
\multirow{3}{*}{Encoder} & \multicolumn{6}{c}{p-value} \\ \cline{2-7} 
 & \multicolumn{3}{c|}{Original} & \multicolumn{3}{c}{FTAL} \\ \cline{2-7} 
 & \multicolumn{1}{c|}{Victim A} & \multicolumn{1}{c|}{Victim B} & \multicolumn{1}{c|}{Victim C} & \multicolumn{1}{c|}{Victim A} & \multicolumn{1}{c|}{Victim B} & Victim C \\ \hline
$F_v$ & \multicolumn{1}{c|}{0} & \multicolumn{1}{c|}{0} & \multicolumn{1}{c|}{0} & \multicolumn{1}{c|}{$10^{-14}$} & \multicolumn{1}{c|}{$10^{-14}$} & $10^{-12}$ \\
$F_p$ & \multicolumn{1}{c|}{$10^{-14}$} & \multicolumn{1}{c|}{$10^{-12}$} & \multicolumn{1}{c|}{$10^{-5}$} & \multicolumn{1}{c|}{$10^{-10}$} & \multicolumn{1}{c|}{$10^{-9}$} & $10^{-4}$ \\ \hline
\multirow{2}{*}{} & \multicolumn{3}{c|}{RTAL} & \multicolumn{3}{c}{Pruning} \\ \cline{2-7} 
 & \multicolumn{1}{c|}{Victim A} & \multicolumn{1}{c|}{Victim B} & \multicolumn{1}{c|}{Victim C} & \multicolumn{1}{c|}{Victim A} & \multicolumn{1}{c|}{Victim B} & Victim C \\ \hline
$F_v$ & \multicolumn{1}{c|}{$10^{-12}$} & \multicolumn{1}{c|}{$10^{-12}$} & \multicolumn{1}{c|}{$10^{-11}$} & \multicolumn{1}{c|}{$10^{-16}$} & \multicolumn{1}{c|}{$10^{-15}$} & $10^{-11}$ \\
$F_p$ & \multicolumn{1}{c|}{$10^{-6}$} & \multicolumn{1}{c|}{$10^{-5}$} & \multicolumn{1}{c|}{$10^{-2}$} & \multicolumn{1}{c|}{$10^{-13}$} & \multicolumn{1}{c|}{$10^{-10}$} & $10^{-4}$ \\ \hline
\multirow{2}{*}{} & \multicolumn{3}{c|}{Noising} & \multicolumn{3}{c}{Shuffle} \\ \cline{2-7} 
 & \multicolumn{1}{c|}{Victim A} & \multicolumn{1}{c|}{Victim B} & \multicolumn{1}{c|}{Victim C} & \multicolumn{1}{c|}{Victim A} & \multicolumn{1}{c|}{Victim B} & Victim C \\ \hline
$F_v$ & \multicolumn{1}{c|}{$10^{-16}$} & \multicolumn{1}{c|}{$10^{-15}$} & \multicolumn{1}{c|}{$10^{-12}$} & \multicolumn{1}{c|}{$10^{-10}$} & \multicolumn{1}{c|}{$10^{-10}$} & $10^{-9}$ \\
$F_p$ & \multicolumn{1}{c|}{$10^{-11}$} & \multicolumn{1}{c|}{$10^{-9}$} & \multicolumn{1}{c|}{$10^{-4}$} & \multicolumn{1}{c|}{$10^{-7}$} & \multicolumn{1}{c|}{$10^{-5}$} & $10^{-3}$ \\ \hline\hline
\end{tabular}%
}
\end{table}

\subsection{Ablation Study}

 Given learned victim encoders, piracy encoders and independent encoders, the purpose of this experiment is to evaluate the impact of multiple factors used in learning the fingerprint, and number of query data in fingerprint verification.

\subsubsection{Impact of Training Dataset}
We evaluate the impact of training and query dataset on $ebr$, and the other settings are unchanged as in Table III. We use three datasets, ImageNet, STL-10 and CIFAR-10, to learn the fingerprint and test victim A. To fully observe the tendency, we record each $ebr$ corresponding to each number of epoch during learning. Three curves corresponding to the three datasets are plotted in Fig.11 (a). The trend of the three curves are consistent and $ebr$ stably decrease with the number of epoch increasing. But there exists discrepancy among the three curves and STL-10 yields the best performance. The result indicates the importance of cover image for the StegGuard.

\subsubsection{Impact of Number of Training Data}
We evaluate the impact of number of training data on $ebr$, and the other settings are unchanged as in Table III. We change the number of training data to learn the fingerprint, and test victim A,B,C. The experimental results shown in Fig. 11(b) indicate that the performance of the fingerprint degrades when the number of training data is less than half of the default number, but even when the number of training data is reduced to 1/5, only 1K training samples, our method can still effectively verify the ownership of the encoder. 

\textbf{\begin{figure}[!ht]
\begin{center}
\includegraphics[width=\linewidth]{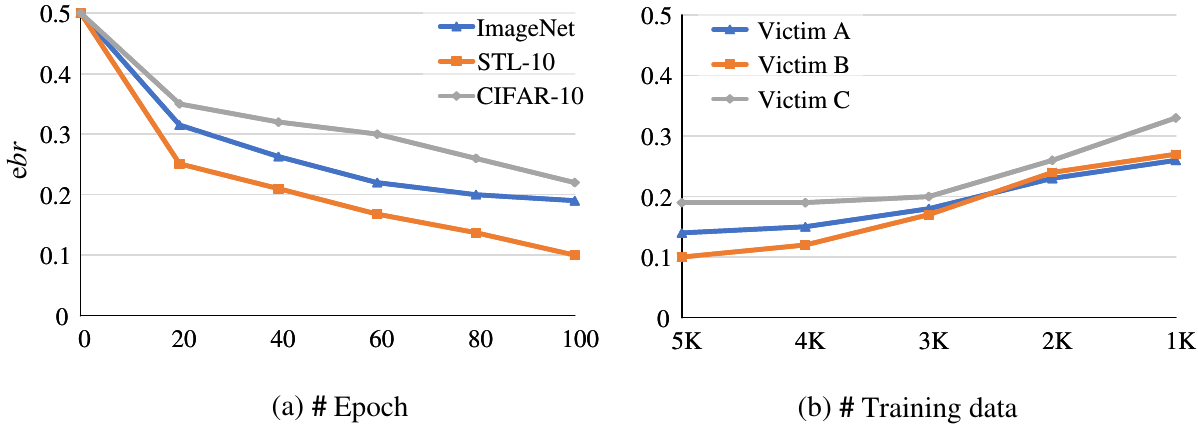}
\caption{The impact of different training factor settings during the fingerprint learning phase.}
\end{center}
\end{figure}}

\subsubsection{Impact of Secrets Length}
We evaluate the impact of secrets length on $ebr$, and the other settings are unchanged as in Table III. We change secrets length to learn the fingerprint, and test victim A,B,C. If the secrets is too short, performance is not ensured. Conversely, if the secrets is too long, the feasibility of secrets embedding and extraction is weakened. Thus, it is necessary to choose an appropriate secrets length to both ensure the performance and enhance the feasibility. As a too short of secrets length is adverse for performance, we start the test from 16bits and end with 512bits. As shown in Fig. 12, when the secrets length comes at 128bits, $ebr$ has a clear increasing. In order to balance between performance and feasibility, 64bits is considered as the best option. 
\textbf{\begin{figure}[!ht]
\begin{center}
\includegraphics[width=0.5\linewidth]{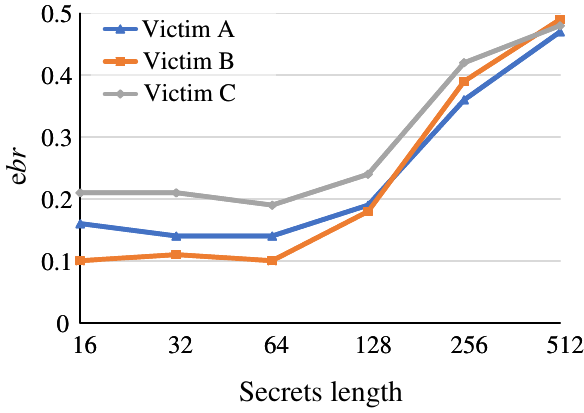}
\caption{The impact of different secrets length during the fingerprint learning phase.}
\end{center}
\end{figure}}

\subsubsection{Impact of Number of Query Data}
We evaluate the impact of number of query data on p-value, and the other settings are unchanged as in Table III. We change number of query data to test 20 piracy encoders and 20 independent encoders produced from victim A by using the same ways as stated in section IV.B and IV.C. StegGuard has an advantage over other works that rely on the distribution similarity of output embeddings as the encoder fingerprint. StegGuard does not require a large number of query images to verify the encoder copyright based on their distribution. We only need to embed different secrets into a small number of query images and feed the stego images into the encoder to obtain the output embeddings. Then the secrets can be extracted through the secrets extractor and compared with the embedded secrets. Thus the ownership of the encoder can be verified by calculating $ebr$. Therefore, we compare with DI \cite{paper29}, and the experimental results are shown in Fig. 13. When the number of query data decreases gradually, the p-value of DI has a large increase, while StegGuard still remains stable even when the number of query data decreases to 0.1K.

\textbf{\begin{figure}[!ht]
\begin{center}
\includegraphics[width=0.6\linewidth]{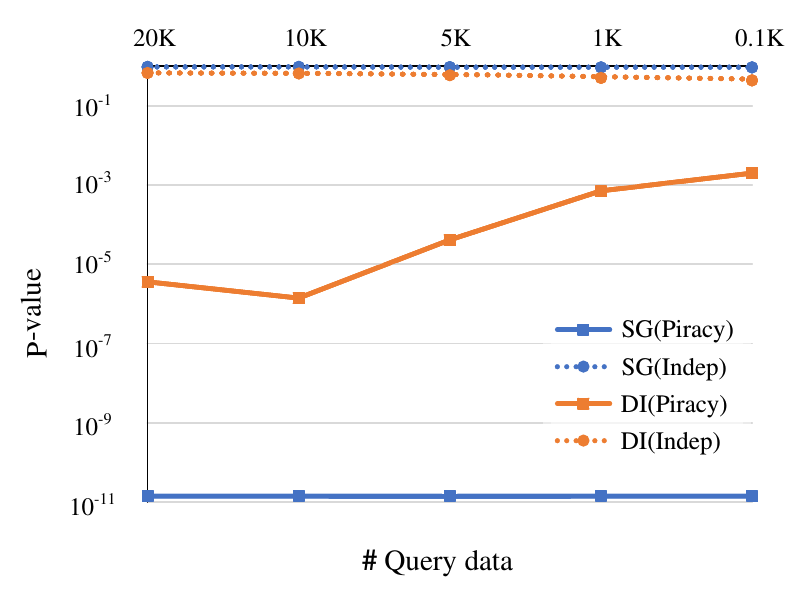}
\caption{Comparison of StegGuard and DI in terms of the number of query data.}
\end{center}
\end{figure}}

\subsubsection{Impact of FcaEmb}
We evaluate the impact of FcaEmd on $ebr$ and visual quality of stego image, and the other settings are unchanged as in Table III. We learn the fingerprint that secrets embeder contains no FcaEmb block, and test victim A,B,C. In section III, we introduce the FcaEmb block into the secrets embeder, which can fully use the medium and high-frequency bands. When the encoder owner attempts to embed the secret into the query image, adaptively embedding the secrets into suitable frequency bands via the FcaEmb block can reduce $ebr$ during the secrets extracting process, and it can also make the stego images visually imperceptible. In other words, FcaEmb enables copyright owners to verify piracy encoders with higher confidence. Besides, after adding the FcaEmb block, the noise in the low-frequency part of the stego image is significantly reduced. This results in weaker texture perturbations in the image and the results are shown in Fig. 14.

\textbf{\begin{figure}[!ht]
\begin{center}
\vspace{-0.3cm}
\includegraphics[width=\linewidth]{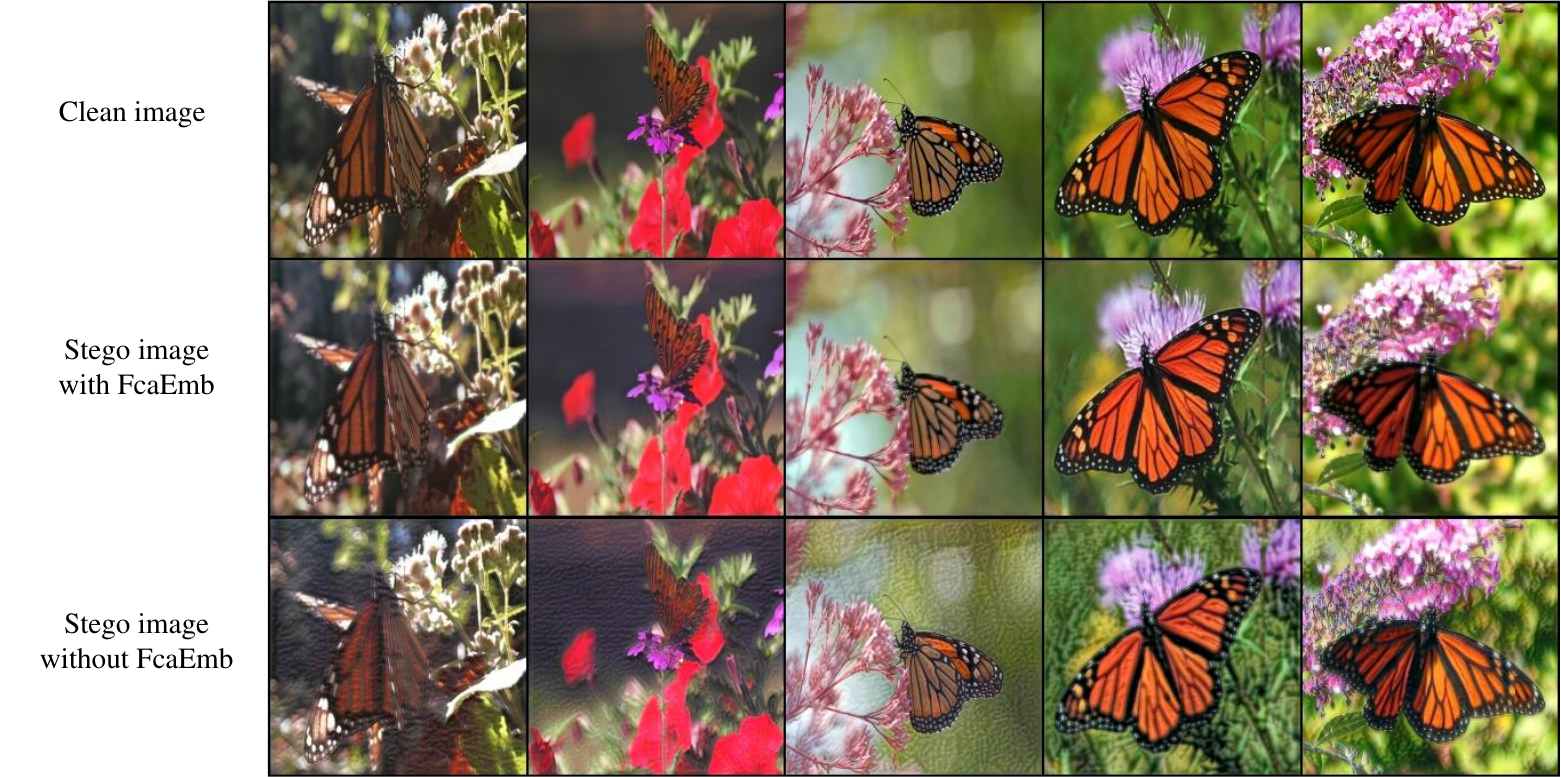}
\caption{The visualization of the clean image(the first row), stego image produced by the secrets embeder with FcaEmb(the second row) and the stego image produced by the secrets embeder without FcaEmb(the third row).}
\end{center}
\end{figure}}

In addition, we also quantitatively compare $ebr$, PSNR and SSIM under different victim encoders with or without adding the FcaEmb block. The experimental results are shown in Table VI. When the FcaEmb block is added into the secrets embeder, $ebr$ will reduce to a certain extent, which is more conducive to the copyright owner determining whether the encoder is a piracy encoder. And after adding the FcaEmb block, the PSNR and SSIM of the stego images are also reduced, which indicates that the stego images have better visual quality and are less likely to be detected by the adversary.

\begin{table}[!ht]
\caption{The impact of FcaEmb block}
\centering 
\resizebox{\columnwidth}{!}{%
\begin{tabular}{c|c|c|c|c}
\hline\hline
Victim Encoder & Module & $ebr$ & PSNR & SSIM \\ \hline
\multirow{2}{*}{Victim A} & w.o.FcaEmb & 0.16 & 28.2 & 0.92 \\
 & w.FcaEmb & \textbf{0.14} & \textbf{29.8} & \textbf{0.95} \\ \hline
\multirow{2}{*}{Vcitim B} & w.o.FcaEmb & 0.11 & 32.4 & 0.90 \\
 & w.FcaEmb & \textbf{0.10} & \textbf{34.9} & \textbf{0.92} \\ \hline
\multirow{2}{*}{Victim C} & w.o.FcaEmb & 0.19 & 30.3 & 0.92 \\
 & w.FcaEmb & \textbf{0.18} & \textbf{32.1} & \textbf{0.94} \\ \hline\hline
\end{tabular}%
}
\end{table}

\subsection{Explanation of StegGuard}
In order to further explain the effectiveness of StegGuard, we introduce GradCAM \cite{paper45} technique to visualize the stego image. GradCAM as a visualization tool is often used to explore the explainability of classification networks. The attention heat map will be generated by back-propagate the gradient through the output prediction results and cross entropy of ground truth. Then rendering the model's attention to the image on the input image. The explanation of the classification results of this model will be obtained by overlaying the heat map onto the original image and brighter colored areas indicate where the model pays more attention. Since StegGuard is the encoder-oriented framework, which is different from the classification model, so that we have made certain changes to GradCAM. In detail, we input the clean image and the stego image at the same time, and obtain the clean embedding and the embedding carrying the secrets information. Then, we take the clean embedding as the benchmark, and find the distance between the clean embedding and the embedding for stego image, and back-propagate to the input image through the gradient. The encoder's attention to the stego image can be observed.

\textbf{\begin{figure}[!ht]
\begin{center}
\includegraphics[width=0.95\linewidth]{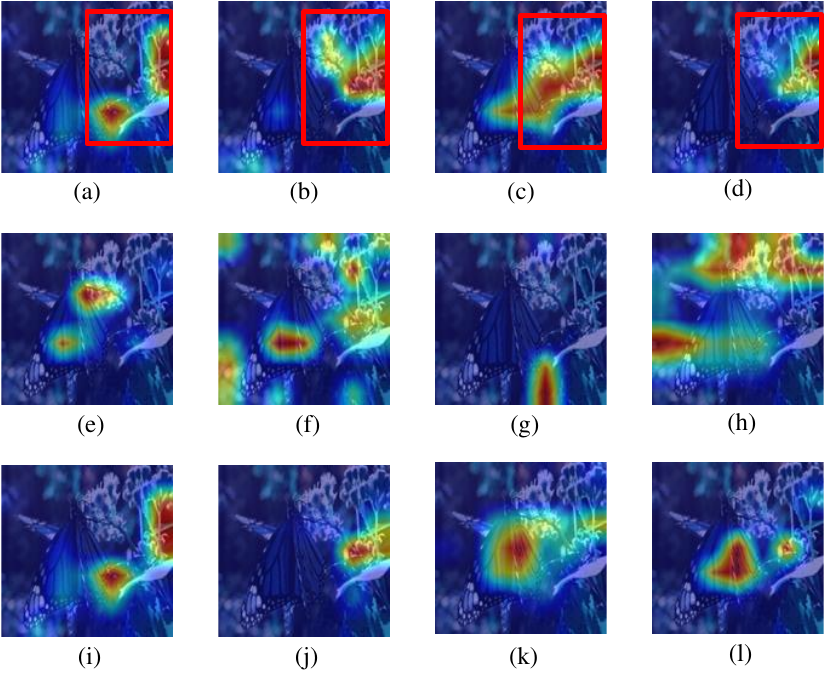}
\caption{The explanation of the StegGuard. (a) is the visualization of the victim encoder; (b), (c) and (d) are the visualization of the piracy encoders; (e), (f), (g) and (h) are the visualization of independent encoders; (i) (j) (k) (l) are the visualization of the victim encoder fine-tuning with different epochs, 25,50,70,100.}
\end{center}
\end{figure}}

The results are shown in Fig. 15, where (a) is the heat map of the victim encoder, and its red area is concentrated in the upper right corner of the image. Besides, (b), (c) and (d) are the heat map of the piracy encoders, and the position of the red area is similar to (a). On the contrary, (e), (f), (g) and (h) show the independent encoders which are pre-trained independently, where the positions of the red regions have a large difference from the victim encoder. (i), (j), (k) and (l) show the results of different epochs, 25,50,70,100, during the process of fine-tuning the victim encoder. When we fine-tune a few epochs, the position of the red regions is still approximately the position of the victim encoder, while the position of the red region is gradually offset as the number increases. The experimental results show that similar encoders have similar feature extraction capabilities for stego images, while the encoders with large differences cannot accurately extract the features in the stego images.

\section{Conclusion}

In this work, we propose StegGuard, a novel fingerprint mechanism for protecting the IP of SSL pre-trained encoder. We argue that an effective fingerprint for a pre-trained encoder can be a pair of secrets embeder and extractor in that how an embeder embeds secrets into images and how an extractor extracts the secrets from embeddings with a tolerable error after the secrets are subjected to the encoder’s transformation relies on the unique transform function of each individual encoder. Moreover, due to the piracy’s dependency on the victim, their transform functions are similar. Based on these, we facilitate StegGuard by joint using the learned fingerprint and the suspect encoder. In terms of $ebr$, the ownership of the suspect encoder is verified. Also, to boost the performance of StegGuard, we introduce the FcaEmb block into the secrets embeder. In the future, we will attempt to extend the proposed idea into multi-modal cases and investigate IP protection for multi-modal models.

\section{Acknowledgements}

This work is supported by CCF-AFSG Research Fund under Grant RF20220019 and the National Natural Science Foundation of China (No.61902235).

\vspace{11pt}

\vfill

\end{document}